\def\ifhtx{\iffalse}    
\ifhtx
\else
\documentclass[11pt,notitlepage,onecolumn]{ivoa}
\fi

\def\SVN$#1: #2 ${\expandafter\def\csname SVN#1\endcsname{#2}}
\SVN$Revision: 2754 $
\SVN$Date: 2014-10-27 12:15:25 +0100 (lun, 27 ott 2014) $
\SVN$HeadURL: https://volute.g-vo.org/svn/trunk/projects/std-vounits/VOUnits.tex $

\usepackage{natbib} 

\usepackage{prettyref} 
\newrefformat{sec}{Sect.~\ref{#1}}
\newrefformat{appx}{Appx.~\ref{#1}}
\newrefformat{fig}{Fig.~\ref{#1}}
\newrefformat{tab}{Table~\ref{#1}}
\usepackage{varioref}
\newrefformat{tabx}{Table~\vref{#1}}

\usepackage{array}
\newcolumntype{L}[1]{>{\raggedright\let\newline\\\arraybackslash\hspace{0pt}}p{#1}}

\makeatletter
\def\units{\@ifstar{\let\un@tsspace\relax    \un@ts}%
                   {\let\un@tsspace\thinspace\un@ts}}
\newcommand{\un@ts}[1]{{\let~\thinspace
  \ifmmode
    \un@tsspace\mathrm{#1}%
  \else
    \nobreak$\un@tsspace\mathrm{#1}$%
  \fi}}

\newcommand*\hex[1]{\uppercase{#1}${}_{16}$}

\usepackage{verbatim} 
\def\verbatim@font{\fontsize{9}{11}\selectfont\ttfamily}

\newcommand*\norm[1]{\textbf{\color{ivoacolor}#1}}

\makeatother

\def\eg{e.g.,~}

\ifx\pdftexversion\undefined
  \usepackage[dvips]{graphicx}
  \DeclareGraphicsExtensions{.eps,.ps}
  \usepackage[ps2pdf]{hyperref}

\else
  \usepackage[pdftex]{graphicx} 
  \usepackage{epstopdf}         
  \makeatletter
  \g@addto@macro\Gin@extensions{,.ps}
  \@namedef{Gin@rule@.ps}#1{{pdf}{.pdf}{`ps2pdf #1}}
  \makeatother
 \usepackage[pdftex,bookmarks=true,bookmarksnumbered=true,hypertexnames=false,breaklinks=true,%
  colorlinks,allcolors=ivoacolor]{hyperref}
\fi
\usepackage[final]{pdfpages}
\title{Units in the VO}
\date{1.0}

\ivoatype{IVOA Recommendation}

\version{REC-1.0}
\author{Markus Demleitner\\
S\'{e}bastien Derri\`ere\\
Norman Gray\\
Mireille Louys\\
Fran\c{c}ois Ochsenbein}
\editor{S\'{e}bastien Derri\`{e}re and Norman Gray}

\urlthisversion{\footnotesize{\url{http://www.ivoa.net/Documents/VOUnits/20140523/}}}
\urllastversion{\footnotesize{\url{http://www.ivoa.net/Documents/VOUnits/}}}
\previousversion{\footnotesize{\url{http://www.ivoa.net/Documents/VOUnits/20140513/}}}

\definecolor{orange}{rgb}{0.5,0.3,0.0}
\newcommand{\unit}[1]{\texttt{\small\color{orange}#1}}
\usepackage[T1]{fontenc}
\usepackage{longtable}
\usepackage{multirow}

\usepackage{upgreek}
\def\micro{{\ensuremath \upmu}}

\newcommand{\brown}{\textcolor[rgb]{0.50,0.10,0.10}}

\begin{document}
\maketitle 
\thispagestyle{empty}
\begingroup
\vfill
\hbox to \textwidth{\hfil\tiny code.google.com/p/volute, rev\SVNRevision, \SVNDate}
\endgroup
\newpage
\tableofcontents 
\newpage
\listoftables
\newpage
\section*{Abstract}
This document describes a recommended syntax for writing the string
representation of unit labels (`VOUnits').  In addition, it describes
a set of recognised and deprecated units, which is as far as possible
consistent with other relevant standards (BIPM, ISO/IEC and the IAU).

The intention is that units written to conform to this specification
will likely also be parsable by other well-known parsers.  To this
end, we include machine-readable grammars for other units syntaxes.

\section*{Status of this document}

This document has been produced by the IVOA Semantics Working Group.
It has been reviewed by IVOA Members and other interested parties, and
has been endorsed by the IVOA Executive Committee as an IVOA
Recommendation. It is a stable document and may be used as reference
material or cited as a normative reference from another
document. IVOA's role in making the Recommendation is to draw
attention to the specification and to promote its widespread
deployment. This enhances the functionality and interoperability
inside the Astronomical Community.

The place for discussions related to this document is the
Semantics IVOA mailing list {\tt semantics\@@ivoa.net}.

A list of current IVOA recommendations and other technical documents can be found at
\url{http://www.ivoa.net/Documents/}.

\subsection*{Note on conformance}

Text within the following document is classified as either
`normative' or `informative'.

\textbf{Normative} text means information that is required
to implement the Recommendation; an implementation of this
Recommendation is conformant if it abides by all the prescriptions
contained in normative text.  \textbf{Informative} text is
information provided to clarify or illustrate a requirement but which
is not required for conformance.

The sections and subsections of this Recommendation are labeled,
after the section heading, to specify whether they are normative or
informative.  If a subsection is not labeled, it has the same
normativity as its parent section.  References are normative if they
are referred to within normative text.

When found within normative sections, the key words
\norm{must},
\norm{must not},
\norm{required},
\norm{shall},
\norm{shall not},
\norm{should},
\norm{should not},
\norm{recommended},
\norm{may},
\norm{optional},
thus formatted, are to be interpreted as described in RFC 2119
\citep{std:rfc2119}.

\section*{Acknowledgements}

We thank all those participants in IVOA and EuroVO workshops who have
contributed by exposing use cases and providing comments during this
document's long development.  These include
Bob Hanisch,
Rick Hessman,
Paddy Leahy,
Jeff Lusted,
Jonathan McDowell,
Marco Molinaro,
Pedro Osuna,
Anita Richards,
Bruno Rino,
Arnold Rots,
Jesus Salgado,
Mark Taylor,
Brian Thomas,
and recent contributors on the DM and Semantics forums.

\section{Introduction (informative)}
\label{sec:intro}

This document describes a standardised use of units in the VO
(hereafter simply `VOUnits').  It aims to describe a syntax for unit
strings which is as far as possible in the intersection of existing
syntaxes, and to list a set of `known units' which is
the union of the `known units' of those standards.
We \emph{recommend}, therefore, that applications which write out
units should do so using \emph{only} the VOUnits syntax, and that
applications reading units should be able to read \emph{at least} the
VOUnits syntax, plus all of the units of \prettyref{sec:knownunits}.
It is not, however, quite possible for VOUnits to be in the
intersection of existing syntaxes; there is futher discussion of this
point in \prettyref{sec:deviations}.

We also provide, for information, a set of self- and mutually-consistent
machine-readable grammars for all of the syntaxes discussed.

The introduction gives the motivation for
this proposal in the context of the VO architecture, from the legacy 
metadata available in the resource layer, to the requirements of the various 
VO protocols and standards and applications.

This document is organised as follows. \prettyref{sec:proposal}
details the proposal for VOUnits. \prettyref{sec:useCase} lists some
use cases and reference implementations.  In \prettyref{appx:current},
there is a brief review of current practices in the description and
usage of units; in \prettyref{appx:comparisons} there is a detailed
discussion of the differences between the various syntaxes; and
in \prettyref{appx:grammar} there are formal (yacc-style) grammars for
the four syntaxes discussed.

The normative content of this document is \prettyref{sec:proposal} and \prettyref{appx:vougrammar}.

\subsection{Units in the VO Architecture}

\renewcommand{\topfraction}{.85}
\renewcommand{\bottomfraction}{.7}
\renewcommand{\textfraction}{.15}
\renewcommand{\floatpagefraction}{.66}

\begin{figure}
  \centerline{\includegraphics[width=0.9\textwidth]{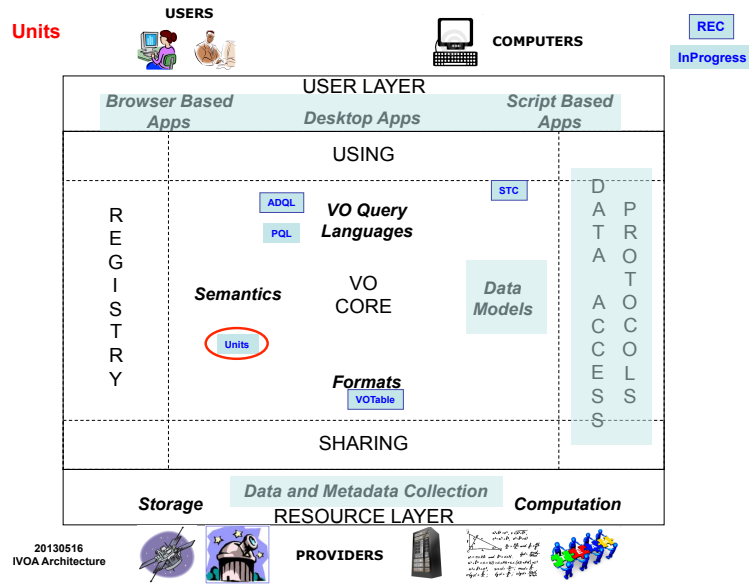}}
  \caption{Units is a core building block in the VO.  Most parts of the
  architecture rely on it: the User Layer with tools and clients, the
  Resource Layer with data.  Protocols, registries entries, and
  data models also re-use these Units definitions.}
  \label{fig:architecture}
\end{figure}

Generally, every quantity provided in astronomy has a unit attached to
its value or is unitless (\eg a ratio, or a numerical multiplier).  

Units lie at the core of the VO architecture, as can be seen in \prettyref{fig:architecture}.
Most of the existing data and metadata collections accessible in the resource
layer have some legacy units, which are mandatory for any scientific use of
the corresponding data.  Units can be embedded in data (\eg FITS headers) or be
implied by convention and/or (preferably) specified in metadata.

Units also appear in the VOTable format \citep{ochsenbein11}, through the use
of a {\tt unit} attribute that can be used in the {\tt FIELD}, {\tt PARAM} and {\tt INFO} 
elements. Because of the widespread dependency of many other VO standards on VOTable,
these standards inherit a dependency on Units.

The Units also appear in many Data Models, through the use of dedicated elements in
the models and schemas.
At present, each VO standard either refers to some external reference document, or 
provides explicit examples of the Units to be used in its scope, on a case-by-case
basis.

The registry records can also contain units, for the description of table metadata.
The definition of VO Data Access protocols uses units by specifying in which units the input
parameters have to be expressed, or by restricting the possible units in which some 
output must be returned.

And last but not least, tools can interpret units, for example to display
heterogeneous data in a single diagram by applying conversions to a reference 
unit on each axis.

\subsection{Adopted terms and notations\label{sec:notations}}

Discussions about units often suffer from misunderstandings arising from cultural
differences or ambiguities in the adopted vocabulary. For the sake of clarity, in this 
document, the following concepts are used:

A \textbf{quantity} is the combination of a (numerical) {\em
value}, measured for a {\em concept} and expressed in terms of a given
{\em unit}; there may be other structure to a quantity, such as
uncertainty or even provenance. 
In the VO context, the nature of the concept can be expressed with a UCD or a utype. This document does not address the full issue of
representing quantities, but focusses on the {\em unit} part.

A \textbf{unit} can be expressed in various forms: in natural language
(\eg \emph{metres per second squared}), with a combination of symbols
with typographic conventions (\eg m s$^{-2}$), or by a simplified text
label (\eg \unit{m.s-2}). VOUnit deals with the label form, which is
easier to standardize, parse and exchange. A VOUnit corresponds in the
most general case to a combination of several (possibly prefixed)
symbols with mathematical operations expressed in a controlled syntax.

A \textbf{unit} consists of a sequence of \textbf{unit components},
each of which represents a \textbf{base unit}, possibly modified by a
multiplicative \textbf{prefix} (of one or two characters), and raised
to an integer or rational power.  The whole unit may (in some
syntaxes) be prefixed by a numerical \textbf{scale-factor}.

Each of the \textbf{base units} (for example, the metre) is
represented by a \textbf{base symbol} (for example \unit{m}).  Each
syntax has a number of \textbf{known units}
(\prettyref{sec:knownunits}), for each one of which there is at least
one symbol which identifies only that unit.

A \textbf{symbol} is either a base symbol or a base symbol with a
scaling prefix.

For example, in the unit of \unit{1.663e-1mm.s**-1}, the scale-factor
is $1.663\times10^{-1}$, the two unit-components are \texttt{mm}
and \texttt{s**-1}; the first symbol has base symbol \texttt{m} and
prefix \texttt{m} (for `milli'), and the second has base
symbol \texttt{s}, no prefix, and the power~$-1$.


\subsection{Purpose of this document}
\label{sec:purpose}

The purpose of this document is to provide a reference specification of how
to write VOUnits, in order to maximize interoperability within the VO;
the intention is that VOUnit strings should be reliably
parsable by humans \emph{and} computers, with a single interpretation.
This is broadly the case for the other existing
unit-string syntaxes, although there are some slight ambiguities in
the specifications of these syntaxes (cf \prettyref{appx:grammar}).
We therefore include a set of self- and mutually-consistent
machine-readable grammars for all of the syntaxes discussed.

The unit syntax(es) described here are intended to be human-readable,
to the extent that, for example, a string such as \unit{mm.s**-2} is
human-readable (without this restriction, we could easily define a
much more regular machine-to-machine grammar).  Having an explicit
unit-string grammar means that data providers can write human-readable
strings in the confidence that the result will \emph{additionally} be
machine-readable in a reliable and checkable way.  Or, where a string
is not fully machine readable (because a data provider needs to use a
custom unit such as 'jupiterMass'; see \prettyref{sec:quoting}), that the
string is at least partially machine readable, and that that partial
readability is non-ambiguous.

We aim not to reinvent the wheel, and to be as compliant as possible with
legacy metadata in major archives, and astronomers' habits.

In particular:
\begin{itemize}
\item We describe (\prettyref{appx:current}) a number of existing unit
  syntaxes, and mention some ambiguities in their
  definition. Application authors should expect to encounter each of
  the syntaxes mentioned in this document (FITS, OGIP and CDS); all of
  these are broadly endorsed by this specification.
\item In addition to the unit syntaxes described above, there are
  multiple specifications of base and known units
  (we refer, in particular, to 
  specifications from BIPM, ISO/IEC and the IAU);
  these are broadly, but not completely, mutually consistent.
\item Where there are some ambiguities in, or contradictions between,
  these various specifications, we recommend that application authors should
  resolve them as indicated in this specification.  
\item This document defines a syntax, called `VOUnits', which is as
  far as is feasible in the intersection of the three existing
  syntaxes, and which we recommend that applications should use when
  writing unit strings.  This aim is not quite possible in fact, and
  the extensions to it, and the mild deviations from it, are discussed
  below in \prettyref{sec:proposal} and \prettyref{appx:grammar};
  there is a summary of the various units
  in \prettyref{tabx:knownunits}.
\end{itemize}


\subsection{What this document will not do}
\label{sec:outofscope}

This Recommendation does \textbf{not} prescribe what units data
providers employ, except to the extent that we avoid giving a standard
interpretation for a unit in some cases (for example we do not
acknowledge the degree celsius or the century as units).  Since we do
not forbid `unrecognised' units, this need not restrict data providers.
Nor do we demand that a given quantity be expressed in a
unique way (\eg all distances in \unit{m}).  So long as data is
labelled in a recognised system, a translation layer can be
provided. Data providers can customise the translation tools if
required. Depending on preference and the operations required, the
user may have a choice of units for his or her query and for the
result.  In particular, the Recommendation does not require that only
recognised units are used.  While it is obviously desirable for data
providers to use recognised and non-deprecated units where possible,
there are occasions when this is unnecessary or undesirable.

This Recommendation does not discuss \emph{quantities} at all.  That
is, we do not discuss the combination of number and unit which refers
to a particular physical measurement, such as `2$\mathrm m\,\mathrm
s^{-1}$'.  Though this might appear to be a trivial extension, it
raises questions of the representation of decimal numbers, the
representation of uncertainties, questions of unit conversion, and
other data-modelling imponderables which have in the past, possibly
surprisingly, generated a great deal of discussion within the
IVOA without, so far, a generally acceptable resolution.

This Recommendation describes only isolated units, and not arrays,
records or other combinations of units.  Several VO protocols require
embedding complex objects into result tables, and give string
serializations for those: geometries in TAP results are the most
common example.  This specification does not cover this situation,
although we hope that where individual unit strings are required in
such instances, their syntax will conform to, or include, this
specification by reference.

In general, this Recommendation is concerned almost exclusively with
the syntactic question of what is and is not a valid unit string,
leaving most questions of interpretation or enforcement to a higher layer in an
application stack.  Specifically:
\begin{itemize}
\item The specification does not forbid `unknown' units.  An
implementation of this specification should be able to recognise, and
communicate, that a unit is unknown, but it is not required to reject
a unit string on the grounds that it is unrecognised.

\item Similarly, although \prettyref{tabx:knownunits} forbids some
units from having SI prefixes, a VOUnit implementation should not
itself reject a unit string which incorrectly includes a prefix, but
should instead just make available the information that this has been
detected, and that it is deprecated.

\item The list of known units in \prettyref{sec:knownunits} is not
specific about the precise definitions of the units in question; for
example, it refers to the `second' without distinguishing between the
various possible definitions that the second may have.
See that section for further discussion.

\item This Recommendation does not specify how an application should
compare units for equivalence; for example, an application may or may
not wish to deem \unit{m/s} and \unit{km/s} to be `equivalent'.
This Recommendation, similarly, does not specify how to compare units
with scale-factors (cf \prettyref{sec:scaleFactors}).
\end{itemize}

\section{The VOUnits syntax (normative)\label{sec:proposal}}

The rules for VOUnits are defined in this section.
Various aspects are addressed:
\begin{itemize}
\item how the labels are encoded;
\item what base symbols are allowed and how they are spelled;
\item what prefixes are allowed and how they are used;
\item how symbols are combined.
\end{itemize}
A formal grammar summarizing these conventions is given
in \prettyref{appx:vougrammar}.

The text below is expected to be compatible with the prescriptions
of the SI standard \citep{si-brochure}, except where noted.

\subsection{String representation and encoding\label{sec:encoding}}

VOUnits may occur in legacy contexts, in which the presence of
non-ASCII characters may cause considerable technical inconvenience
(for example FITS cards).  There are only a few non-ASCII characters
which we might wish to include in unit strings (for example \AA\
or \micro), and we can find substitutes for these sufficiently easily, that we
feel there is little real benefit in permitting non-ASCII characters
in VOUnit strings.

All the VOUnit characters in the specification below are printable ASCII
characters (that is, in the range hexadecimal 20 to 7E); any
extensions to this standard \norm{should} be restricted to this same range.

All VOUnit strings \norm{must} be regarded as case-sensitive (the
strings in the other syntaxes are also case-sensitive).

\subsection{Parsing unit strings -- overview\label{sec:parsing-components}}

The unit strings \unit{unknown} and \unit{UNKNOWN} (that is, in
all-lowercase or all-uppercase) are reserved for cases
when the unit is unknown; that is, it is known that there should be a
unit, but the unit string has been lost or not been specified.  These
strings are not, however, part of the list of known units or the VOUnits grammar,
and applications \norm{must} check for their presence before unit parsing.

An empty unit string positively indicates that the corresponding
quantity is dimensionless.  Since an empty string does not conform to
the grammars below, this also \norm{must} be checked for before
unit-parsing starts.

A \textbf{symbol} within a unit-component \norm{should} be parsed as follows:
\begin{enumerate}
\item If it corresponds to a known \textbf{base symbol}, then it
\norm{must} be recognised as such (for example the \unit{Pa} must be
parsed as the known Pascal, and never as the peta-year).

\item If the symbol starts with a multiplicative prefix, then this is
recognised independently of whether the resulting base symbol is a
known or unknown unit -- thus \unit{Mm} and \unit{Mfurlong} are parsed
as millions of metres and furlongs, but note that this implies, for
the sake of consistency, that \unit{furlong} is parsed as the
femto-`urlong'.

\item In the VOUnits syntax (a significant divergence from the other
syntaxes), base symbols \norm{may} be put between single
quotes \unit{'...'} (ASCII character \hex{27}).
Such symbols \norm{must} be parsed as
unrecognised unit symbols which are not further examined.
See \prettyref{sec:quoting} for discussion.
\end{enumerate}

A library which implements this specification \norm{should} be able to
distinguish known and unknown units, and identify deviations from the
restrictions on their use, below.  It \norm{should} be able to
communicate such information to a caller, but it \norm{should not}
unilaterally reject unit strings which use unknown units or use known
units in disapproved ways (of course, a higher-level application is
free to reject unit strings for any reason it pleases).

\subsection{Base units\label{sec:baseUnits}}

There is good agreement for the base symbols across the different schemes
(see \prettyref{tabx:comparUnitBase}).

The VOUnits base symbols are listed in \prettyref{tab:voubase}

\begin{table}[ht]
\begin{center}
\def\arraystretch{1.2}
\begin{tabular}{|rl|rl|rl|rl|}\hline
\unit{m}&(metre)		&\unit{g}&(gram) 	&\unit{J}&(joule)     	&\unit{Wb}&(weber)\\
\unit{s}&(second)		&\unit{rad}&(radian)    &\unit{W}&(watt) 	&\unit{T}&(tesla)\\
\unit{A}&(ampere)		&\unit{sr}&(steradian)  &\unit{C}&(coulomb)	&\unit{H}&(henry)\\
\unit{K}&(kelvin)		&\unit{Hz}&(hertz)      &\unit{V}&(volt) 	&\unit{lm}&(lumen)\\
\unit{mol}&(mole)		&\unit{N}&(newton)      &\unit{S}&(siemens)	&\unit{lx}&(lux)\\
\unit{cd}&(candela)		&\unit{Pa}&(pascal)     &\unit{F}&(farad)	&\unit{Ohm}&(ohm)\\\hline
\end{tabular}
\end{center}
\caption{\label{tab:voubase}VOUnits base units}
\end{table}

For masses, the SI unit is \unit{kg}. However, existing specifications
recommend not using scale-factors with \unit{kg}, but attaching them
only to \unit{g} instead.

Recognising a known unit takes priority over parsing for prefixes.
Thus the string \unit{Pa} represents the Pascal, and not the
peta-year, and the string \unit{mol} will always be the mole, and
never a milli-`ol', for some unknown unit~`ol'.

\subsection{Known units}
\label{sec:knownunits}

In \prettyref{tabx:knownunits}, we indicate the `known units' for each of the
described syntaxes, which go beyond the physically motivated set of
base units.
There are a few units (namely \unit{angstrom} or \unit{Angstrom}, 
\unit{pix} or \unit{pixel}, \unit{ph} or \unit{photon} and \unit{a} or \unit{yr}) for
which there are recognised alternatives in some syntaxes, and in these
cases `p' marks the preferred one.

This list of known units is not specific about the precise definitions
of the units in question; for example, it refers to the `second'
without distinguishing between the various possible definitions that
the second may have: they may be mean-solar or atomic seconds, and be
defined at different points in spacetime.  Generally, when data is
exchanged in those areas where such distinctions matter -- such as
data connected with pulsar timings -- the fine semantic details will
be indicated by the data provider through other mechanisms.  That
said, a VOUnits processor must interpret the symbols
of \prettyref{tabx:knownunits} compatibly with the indicated units:
a \unit{m} is always a metre of one type or another, and may not be
interpreted as, for example, a minute.

\emph{Unrecognised units \norm{should} be accepted by parsers},
as long as they are parsed giving preference to the syntaxes and
prefixes described here.  Thus, for example, the
string \unit{furlong/week} \norm{should} parse successfully (though
perhaps with suitably prominent warnings) as the femto-`urlong' per
week.

The Unity library (\prettyref{sec:libraries}) recognises units with
respect to a subset of the QUDT unit framework~\cite{qudt}, with some
astronomy-specific additions.  This is a particularly comprehensive
collection of units, and we commend it to the IVOA community as
a \emph{lingua franca} for this type of work.

Sections \ref{sec:binary} to \ref{sec:other} below, discussing the set
of known units, are longer than one might expect would be necessary.
Most of the discussion concerns rather arcane edge-cases, or attempts
to reconcile the minor deviations between the relevant existing
standards.  In all cases, we have attempted to be as uninnovative and
unsurprising as possible.

Future versions of this specification may add to the set of known units.

\begin{table}
\hbox to \textwidth{\hss
\catcode`\%=11
\begin{tabular}{rlcccc|rlcccc}
\emph{symbol}&\emph{description}&\emph{fits}&\emph{ogip}&\emph{cds}&\emph{vou}&
\emph{symbol}&\emph{description}&\emph{fits}&\emph{ogip}&\emph{cds}&\emph{vou}\\
A&ampere&s&s&s&s & K&kelvin&s&s&s&s\\
a&julian year&s&&s&s & lm&lumen&s&s&s&s\\
adu&ADU&$\cdot$&&&s & lx&lux&s&s&s&s\\
Angstrom&angstrom&d&&$\cdot$&dp & lyr&light year&$\cdot$&$\cdot$&&s\\
angstrom&angstrom&&$\cdot$&&d & m&meter&s&s&s&s\\
arcmin&arc minute&$\cdot$&$\cdot$&$\cdot$&s & mag&magnitudes&s&$\cdot$&s&s\\
arcsec&arc second&$\cdot$&$\cdot$&s&s & mas&milliarcsecond&$\cdot$&&$\cdot$&$\cdot$\\
AU&astronomical unit&$\cdot$&$\cdot$&$\cdot$&p & min&minute (time)&$\cdot$&$\cdot$&$\cdot$&s\\
au&astronomical unit&&&&$\cdot$ & mol&mole&s&s&s&s\\
Ba&besselian year&&&&d & N&newton&s&s&s&s\\
barn&barn&sd&$\cdot$&s&sd & Ohm&ohm&s&&s&s\\
beam&beam&$\cdot$&&&s & ohm&ohm&&s&&\\
bin&bin&$\cdot$&$\cdot$&&s & Pa&pascal&s&s&s&s\\
bit&bit&s&&s&sb & pc&parsec&s&s&s&s\\
byte&byte&s&$\cdot$&s&sbp & ph&photon&$\cdot$&&&s\\
B&byte&&&&sb & photon&photon&p&$\cdot$&&sp\\
C&coulomb&s&s&s&s & pix&pixel&$\cdot$&&$\cdot$&s\\
cd&candela&s&s&s&s & pixel&pixel&p&$\cdot$&&sp\\
chan&channel&$\cdot$&$\cdot$&&s & R&rayleigh&s&&&s\\
count&number&$\cdot$&$\cdot$&&sp & rad&radian&s&s&s&s\\
Crab&crab&&s&& & Ry&rydberg&$\cdot$&&s&s\\
ct&number&$\cdot$&&$\cdot$&s & s&second (time)&s&s&s&s\\
cy&julian century&$\cdot$&&& & S&siemens&s&s&s&s\\
d&day&$\cdot$&$\cdot$&$\cdot$&s & solLum&luminosity&$\cdot$&&$\cdot$&s\\
dB&decibel&&&&$\cdot$ & solMass&solar mass&$\cdot$&&$\cdot$&s\\
D&debye&$\cdot$&&$\cdot$&s & solRad&solar radius&$\cdot$&&$\cdot$&s\\
deg&degree (angle)&$\cdot$&$\cdot$&$\cdot$&s & sr&steradian&s&s&s&s\\
erg&erg&d&$\cdot$&&sd & T&tesla&s&s&s&s\\
eV&electron volt&s&s&s&s & ta&year tropical&&&&d\\
F&farad&s&s&s&s & u&AMU&$\cdot$&&&s\\
g&gramme&s&s&s&s & V&volt&s&s&s&s\\
G&gauss&sd&$\cdot$&&sd & voxel&voxel&$\cdot$&$\cdot$&&s\\
H&henry&s&s&s&s & W&watt&s&s&s&s\\
h&hour&$\cdot$&$\cdot$&$\cdot$&s & Wb&weber&s&s&s&s\\
Hz&hertz&s&s&s&s & yr&julian year&sp&$\cdot$&sp&sp\\
J&joule&s&s&s&s\\

\end{tabular}
\hss}
\caption[Known units in the various syntaxes]
{\label{tabx:knownunits}Known units in the various syntaxes.
In the table, and for a given syntax, a `$\cdot$' indicates that the unit is recognised,
an~`s' that it is additionally permitted to have SI prefixes,
a~`b' that binary prefixes will be recognised,
and a~`d' that it is recognised but deprecated.
For those units which have alternative symbols for a given unit,
a~`p' indicates the one preferred for output.}
\end{table}

\subsection{Binary units}
\label{sec:binary}

The symbol~`b' is sometimes used for `bits', but this is the SI symbol
for `barn', and this Recommendation aligns with the SI standard in
this respect.  Since the same symbol is sometimes used for `bytes', it
is probably best avoided in any case.

\citet[item 13-9.c]{std:iec80000-13} notes that the term `byte'
`has been used for numbers of bits other than eight' in the past, but
that it should now always be used for eight-bit bytes; we recommend
the same interpretation here.  The same source notes the theoretical
confusion between the symbol \unit{B} for `byte' and for `Bel'.  We
believe it would be perverse in our present context to recommend
against using `B' for byte, and resolve this here
in favour of `byte' by mandating that \unit{B} \norm{must} be parsed
as indicating the `byte', that the \unit{dB} is an
unprefixable special-case unit (as discussed below), and by
implication that the `dB'
\norm{must not} be interpreted as a tenth of a byte.\footnote{We have no
evidence that this has been a common source of confusion within the
IVOA, or indeed anywhere else.}

\subsection{Scale factors\label{sec:scaleFactors}}

Units \norm{may} be prefixed by any of the 20 SI scale-factors, 
and a subset \norm{may} be prefixed by the eight binary scale-factors.
The SI scale-factors -- provided in \prettyref{tab:vouscalefactors}a --
are the same as those of \citet{si-brochure},
of \citet[\S6.5.4]{std:iso80000-1},
and of \citet[Table~5]{pence10}
(see also \prettyref{tabx:comparUnitScale} for further comparisons).
\begin{table}
\def\arraystretch{1.2}
\begin{center}
\def\pfx#1#2{#1, $10^{#2}$}
\begin{tabular}{|rl|rl|}\hline
\unit{da}&\pfx{deca}{1}&
  \unit{d}&\pfx{deci}{-1}\\
\unit{h}&\pfx{hecto}{2}&
  \unit{c}&\pfx{centi}{-2}\\
\unit{k}&\pfx{kilo}{3}&
  \unit{m}&\pfx{milli}{-3}\\
\unit{M}&\pfx{mega}{6}&
  \unit{u}&\pfx{micro}{-6}\\
\unit{G}&\pfx{giga}{9}&
  \unit{n}&\pfx{nano}{-9}\\
\unit{T}&\pfx{tera}{12}&
  \unit{p}&\pfx{pico}{-12}\\
\unit{P}&\pfx{peta}{15}&
  \unit{f}&\pfx{femto}{-15}\\
\unit{E}&\pfx{exa}{18}&
  \unit{a}&\pfx{atto}{-18}\\
\unit{Z}&\pfx{zetta}{21}&
  \unit{z}&\pfx{zepto}{-21}\\
\unit{Y}&\pfx{yotta}{24}&
  \unit{y}&\pfx{yocto}{-24}\\
\hline
\end{tabular}
\qquad
\def\pfx#1#2{#1, $2^{#2}$}
\begin{tabular}{|rl|}\hline
\unit{Ki}&\pfx{kibi}{10}\\
\unit{Mi}&\pfx{mebi}{20}\\
\unit{Gi}&\pfx{gibi}{30}\\
\unit{Ti}&\pfx{tebi}{40}\\
\unit{Pi}&\pfx{pebi}{50}\\
\unit{Ei}&\pfx{exbi}{60}\\
\unit{Zi}&\pfx{zebi}{70}\\
\unit{Yi}&\pfx{yobi}{80}\\
\hline
\end{tabular}
\end{center}
\caption[VOUnits prefixes]{\label{tab:vouscalefactors}VOUnits prefixes:
(a, left) decimal prefixes;
(b, right) binary prefixes}
\end{table}

Writers of unit strings \norm{must not} use compound prefixes (that is,
more than one SI prefix). Prefixes are concatenated to the base
symbol without space, and \norm{must not} be used without a base symbol.

The SI prefixes of \prettyref{tab:vouscalefactors}a \emph{\norm{must}
always refer to multiples of 1000}, even when applied to binary units
such as bit or byte; this follows the stipulations (and clarifying note) of
\citet[\S3.1]{si-brochure}, and of \citet[\S6.5.4]{std:iso80000-1}.
If data providers wish to use multiples of 1024 (ie, $2^{10}$) for
units such as bytes or bits, they \norm{must} use the the binary prefixes
of \citet[\S4]{std:iec80000-13}, reproduced in \prettyref{tab:vouscalefactors}b
(these originated in \citet{std:ieee1541-2002}).

Note 1: the~`s' and~`b' annotations in \prettyref{tabx:knownunits} are
not symmetric: the~`s' annotation indicates that SI prefixes are
permitted in the given syntax, which means that they are also
recognised when preceding unknown units (which have no restrictions on
them); in contrast, binary prefixes are recognised exclusively on
units with a~`b' annotation, which means that they are \emph{not}
recognised with unknown units.  That is, the \unit{Mifurlong} is the
mega-\texttt{ifurlong} (because it starts with the decimal `M' prefix)
and the \unit{Kifurlong} is the unknown unit \texttt{Kifurlong}
(because `K' is not a decimal prefix, and the binary prefix `Ki' is not
recognised when prefixed to an unknown unit).

Note 2: The letter \unit{u} is used instead of the
\micro\ symbol to represent a factor of $10^{-6}$, 
following the character set defined in \prettyref{sec:encoding}.

Note 3: The FITS deprecations in \prettyref{tabx:knownunits} are those
where that standard's Table~4 notes that the unit is ``Deprecated in
IAU Style Manual but still in use.''  The VOUnits
deprecations inherit these, and add (by community acclaim) the
besselian and tropical years.

Note 4: The unit \emph{names} `angstrom' and `ohm' are correctly spelled with
a lowercase initial letter (as required by \citet{std:iso80000-1}), but
in both cases their usual unit \emph{symbol} is a non-ASCII character,
so that their all-ASCII version here must be a word.  Following the
lead of the FITS and CDS standards, and of SI unit-symbols derived from
surnames, and disagreeing with the OGIP standard, we have preferred the
initial-capital symbol in VOUnits (thus `Angstrom' as the unit symbol,
rather than `angstrom').

\subsection{Astronomy symbols}

\prettyref{tabx:comparUnitAstro} lists symbols used in astronomy to
describe times, angles, distances and a few additional quantities.
The subset of these used by this specification are
listed in \prettyref{tab:vouadopted}.

\begin{table}[t]
\begin{center}
\def\arraystretch{1.2}
\begin{tabular}{|rl|rl|rl|}\hline
\unit{min}&(minute of time)	&\unit{deg}&(degree of angle) 	&\unit{Jy}&(jansky) 	\\
\unit{h}&(hour of time)		&\unit{arcmin}&(arcminute)    	&\unit{pc}&(parsec) 	\\
\unit{d}&(day)			&\unit{arcsec}&(arcsecond)  	&\unit{eV}&(electron volt)	\\
\unit{a}, \unit{yr}&(year)	&\unit{mas}&(milliarcsecond)    &\unit{AU}&(astronomical\\
\unit{u}&(atomic mass)		&          &                    &         & unit)\\
\hline
\end{tabular}
\end{center}
\caption{\label{tab:vouadopted}Additional astronomy symbols}
\end{table}

Minutes, hours, and days of time \norm{must} be represented in VOUnits by the
symbols \unit{min}, \unit{h} and \unit{d}; however the \unit{cd} is
the candela, not the centi-day.\footnote{We therefore rule out
interpreting \units{dB/cd} as 0.9\units{mbit/s}.}  The year \norm{may} be expressed by
\unit{yr} (common practice),
or \unit{a},
as recommended by ISO \citep[Annex C]{std:iso80000-3}
and the IAU \citep[Table 6]{wilkins89}.
However peta-year must only be written \unit{Pyr},
to avoid the collision with the pascal, \unit{Pa}.

There are no VOUnit symbols for degrees celsius or century.
Temperatures are expressed in kelvin (\unit{K}),
and a century corresponds to \unit{ha} or \unit{hyr}.
Note that \emph{this is a mild deviation from the SI standard},
which states that the `hectare', with unit symbol \unit{ha},
is a `non-SI unit accepted for use' as a measure of land area~\citep[table~6]{si-brochure},
and which acknowledges neither `a' nor `yr' as a symbol for year.\footnote{If
large telescope arrays feel they must talk of attojoules per
hectare per century, for some reason, they're going to have to be
careful how they do so; it's probably best not to even think about atto-Henrys.}

The astronomical unit \norm{should} be expressed in upper-case, \unit{AU}, in
order to follow legacy practice.  It may also be written \unit{au}, in
the VOUnits syntax, on the ground that it would be perverse to prefer
the atto-atomic-mass to the astronomical unit, in an astronomical unit
specification.
\emph{This is a deviation} from the SI recommendation of
`ua'~\citep[Table 7]{si-brochure}, but conformant with the IAU's
recommendation of `au'~\citep{iau12}.%
\footnote{If you feel a burning desire to write about micro-years or
atto atomic-mass, this document is not the place you need to look
for help.}

Because of the near-degeneracy between the decimal prefixes \texttt{d}
and \texttt{da}, there is an ambiguity when parsing the
unit \unit{dadu} -- is this the deka-\unit{du} or the deci-\unit{adu}?
The only cases where this ambiguity is possible are those involving
known units starting with~`a' (\texttt{da} is unambiguously a
deci-year for the same reason that \texttt{d} is unambiguously a day,
because the presence of a bare unit prefix would be ungrammatical).
We can think of no cases where the prefix is useful enough that
resolving the ambiguity is worth the specification effort, so we deem
the parse of \texttt{da.*} to be \textbf{unspecified}. 
In consequence, data providers \norm{must not} use the \texttt{da}
prefix, and \norm{should not} use the \texttt d prefix (as noted
in \prettyref{sec:other}, the decibel, \unit{dB} is listed as a `known
unit', as opposed to a deci-Bel).

\subsection{Other symbols, and other remarks}
\label{sec:other}

\prettyref{tabx:comparUnitDeprecated} corresponds to Table~7 in the IAU document, and the IAU strongly
recommends no longer using these units. 
Data producers are strongly advised to prefer the equivalent notation using symbols and prefixes listed in 
Tables~\ref{tabx:comparUnitBase}, \ref{tabx:comparUnitScale} and \ref{tabx:comparUnitAstro}. 

However, in order to be compatible with legacy metadata, VOUnit
parsers \norm{should} be able to interpret symbols \unit{angstrom}
or \unit{Angstrom} (for \aa{}ngstr\"om), \unit{barn}, \unit{erg}
and \unit{G} (for gauss).

\prettyref{tabx:comparUnitOther} compares other miscellaneous symbols. 
The last set of VOUnits symbols, derived from this comparison, is in
\prettyref{tab:voumisc}.

\begin{table}[ht]
\begin{center}
\def\arraystretch{1.2}
\begin{tabular}{|l|l|L{3cm}|l|}\hline
\unit{mag} (magnitude)		&\unit{pix}  or \unit{pixel} 	&\unit{solMass} (solar mass)     &\unit{R} (rayleigh) \\
\unit{Ry} (rydberg)		&\unit{voxel}    		&\unit{solLum} (solar luminosity)&\unit{chan} (channel) 	\\
\unit{lyr} (light year)		&\unit{bit}   			&\unit{solRad} (solar radius)	&\unit{bin} \\
\unit{ct} or \unit{count}	&\unit{byte} (8 bits)   	&\unit{Sun} (relative to the Sun, e.g. abundances)&\unit{beam} 	\\
\unit{ph} or \unit{photon} 	&\unit{adu}                     &\unit{D} (Debye)	&\unit{unknown} (\prettyref{sec:parsing-components})\\\hline
\end{tabular}
\end{center}
\caption[Miscellaneous VOUnits]
{\label{tab:voumisc}Miscellaneous VOUnits.}
\end{table}

A few symbols which might theoretically be ambiguous are listed in
\prettyref{tab:ambiguous},
with their consensus VOUnit interpretation.

\begin{table}[bht]
\begin{center}
\begin{tabular}{|r|l|l|}
\hline
\textbf{VOUnit}&\textbf{Correct interpretation}&\textbf{Incorrect}\\
\unit{Pa}&pascal&peta-year\\
\unit{ha}&hecto-year&hectare\\
\unit{cd}&candela&centi-day\\
\unit{dB}&decibel&deci-byte\\
\unit{B}&byte&bel\\
\unit{au}&astronomical unit&atto-atomic-mass\\
\hline
\end{tabular}
\end{center}
\caption{\label{tab:ambiguous}Possibly ambiguous units}
\end{table}

It can be noted that some of the units listed in \prettyref{tabx:comparUnitOther} are 
questionable. They arise in fact from a need to describe quantities, when the only
piece of metadata available is the unit label. Count, photon, pixel, bin, voxel, bit,
byte are concepts, just as apple or banana. The associated quantities could be fully
described with a UCD, a value and a void unit label.
It is possible to count a number of bananas, or to express a distance measured in
bananas, but this does not make a banana a reference unit.

The FITS document provides the most general description of all the compared schemes, 
and VOUnits adopts similar definitions, for the sake of legacy metadata.
The VOUnits symbol for magnitudes is \unit{mag}.
Note that all symbols like \unit{count}, \unit{photon}, \unit{pixel}
are always used in lower case and singular form.

The decibel, \unit{dB} is listed in the SI specification
\citep[Table 8]{si-brochure} amongst a set of `other non-SI units',
and mentioned by \citet[\S0.5]{std:iso80000-3} in a `Remark on
logarithmic quantities'.  The \unit{dB} is included in the list of
`known units' of \prettyref{tabx:knownunits} and so \norm{must} be parsed as a
unit by itself -- as opposed to being parsed as the prefix~`d'
qualifying the unit `Bel' -- and both the decibel and Bel \norm{must
not} be used with other scaling prefixes.

If there is no unit associated with a quantity (for example a quantity
that is a character string, or unitless), data providers \norm{should}
indicate this with an empty string rather than blanks or dashes.

\subsection{Mathematical expressions containing symbols}

\prettyref{tabx:comparUnitCombine} summarizes how, 
in the various existing syntaxes, mathematical operations may
be applied on unit symbols for exponentiation, multiplication,
division, and other computations.

The combination rules are where the largest discrepancies between the
different schemes appear. The FITS document discusses the problem of
trying to best accommodate the existing schemes
\cite[\S4.3.1]{pence10}, without really resolving the problem.
\label{sec:fitsquote}
This and other ambiguities are discussed in the detailed syntaxes of \prettyref{appx:grammar}.

VOUnits follow a subset of the FITS rules,
as summarized in \prettyref{tab:VOUnitCombine}.

\begin{table}
\begin{center}
\def\arraystretch{1.2}
\begin{tabular}{|r|l|}
\hline
\unit{str1.str2} & Multiplication \\
\unit{str1/str2} & Division \\
\unit{str1**expr} & Raised to the power expr \\
\unit{fn(str1)} & Function applied to a unit string\\
\hline
\end{tabular}
\end{center}
 \caption[Combination rules and mathematical expressions for VOUnits]
{\label{tab:VOUnitCombine}Combination rules and mathematical expressions for VOUnits.
See \prettyref{appx:vougrammar} for the complete grammar.}
\end{table}

As illustrated in \prettyref{tab:VOUnitCombine}, units may include a
limited set of functional dependencies on other units.  The set of
functions recognised within VOUnits is the same as the set recommended
by FITS, and listed in \prettyref{tab:functions}.  As with
unrecognised units,
\emph{parsers \norm{should} accept unrecognised functions without error},
even if they deprecate them at some later processing stage.  As
described in \prettyref{sec:quoting}, functions may be quoted to
indicate that they \norm{must not} be interpreted as in this table.
\begin{table}
\begin{center}
\def\arraystretch{1.2}
\begin{tabular}{|r|l|}
\hline
\unit{log(str1)} & Common Logarithm (to base 10) \\
\unit{ln(str1)} & Natural Logarithm \\
\unit{exp(str1)} & Exponential (e$^{\mathrm{str1}}$) \\
\unit{sqrt(str1)} & Square root \\
\hline
\end{tabular}
\end{center}
\caption{\label{tab:functions}Functions of units.}
\end{table}
Note that since functions such as `log' require dimensionless
arguments, when a quantity~$x$ is (for example) represented by numbers
labelled with units \unit{log(Hz)}, that indicates that the numbers
are related to~$x$ by the function
$\log\bigl(x/(\mathrm{1\,Hz})\bigr)$.

%
%
%
%

\subsection{The numerical scale-factor}
\label{sec:scalefactor}

A VOUnits unit string \norm{may} start with a numerical scale-factor
to indicate a derived unit.  For example, the inch might appear as the
unit of \unit{25.4mm}.  See \prettyref{appx:vougrammar} for the syntax
of the VOUnits numerical string.

A data provider may choose to use such a unit in order to represent a
unit which is not listed as one of the VOUnit `known units'.  For
example, given a VOTable column of masses relative to Jupiter's mass,
one might label it as having units of \unit{1.898E27kg} rather than
\unit{'jupiterMass'} (an `unknown unit').
The \emph{advantage} of doing so is that the data consumer can
translate the column data into well-known physical units without further
information, and the data source is thus self-contained.
The \emph{disadvantage} of doing so is (i) that the intention might be
obscured (this is a type of provenance information);
and (ii) that the measurements may be relative to (in this example)
the actual mass of Jupiter rather than merely expressed in those terms,
so that the measurements should change if the actual mass were to be
refined as a result of a recalibration, or if (in the case of a pulsar
period for example) the unit were time-varying.  The data provider retains the
choice of which strategy to take.

A data provider may need to provide further metadata information, to
clarify the meaning of such a unit, or they may judge that the meaning
is adequately clear to the intended audience, without further
complication.  Such further information is out of scope of the VOUnits
Recommendation, in the same way that even `known' units may be
ambiguous in some contexts (cf, \prettyref{sec:knownunits}).

This Recommendation does not prescribe how many significant figures
should be in a scale-factor, nor whether it should be interpreted as
single- or double-precision, nor how units with scale-factors should
be compared for equality.  All of these are implementation choices for
the software which is handling the units.

\subsection{Quoting unknown units\label{sec:quoting}}

The VOUnits syntax permits the use of `unknown units' (that is, units not listed
in \prettyref{tabx:knownunits}).  There need be no syntactic indication that
a unit is `unknown'; this is convenient, but creates some minor
ambiguities.

In the VOUnits syntax, base symbols may be put between single
quotes \unit{'...'} (a significant divergence from the other
syntaxes).  Such symbols \norm{must} be parsed as
unrecognised unit symbols which are not further examined.

This has two consequences.  Firstly, it means that an unknown symbol
which happens to start with an SI prefix is not broken
into a base symbol and prefix: thus \unit{'furlong'} is parsed as
expected, whereas \unit{furlong} would be the femto-`urlong'.
Secondly, a quoted symbol is parsed as an unrecognised unit, even if
it would otherwise indicate a known unit; thus the unit \unit{'m'} is
parsed as an unknown unit `m', and does not indicate the metre.

This facility means that a data provider may label data with units of,
for example, \unit{'martianDay'} or the \unit{'B'}, while still
remaining conformant with the VOUnits Recommendation, and without
risking the leading \texttt{m} being misparsed as an SI prefix, or the
`B' being misparsed as a `byte'.

Quoted units can take prefixes (they are `unknown units', so there are
no restrictions on their usage), so that \unit{m'furlong'} is a
milli-furlong, and \unit{m'm'} is a milli-`m'.  The only permissible
prefixes are those of \prettyref{tab:vouscalefactors}.

\subsection{General rationale (informative)}
\label{sec:rationale}

\subsubsection{Deviations from other syntaxes}
\label{sec:deviations}

The aspiration of the VOUnits work was that the syntax should be as
much as possible in the intersection of the various pre-existing
syntaxes, so that a unit string which conformed to the VOUnits syntax
would be parsable in each of those other syntaxes.  This has not been
possible in fact, for four reasons.
\begin{enumerate}
\item The CDS syntax permits only a dot to indicate a product, and the
OGIP syntax only a star, while FITS permits both.  The VOUnits syntax
uses a dot, so that non-trivial OGIP unit strings are therefore
necessarily invalid VOUnits strings in this one respect.
\item The VOUnits syntax permits (but does not require) a scale-factor
at the beginning of the string, which is not a power of 10.  Only the
CDS syntax permits a similar factor.
See \prettyref{sec:scalefactor} for discussion.
\item Only the VOUnits syntax permits quoted units.
\item Only the VOUnits syntax permits the use of the binary prefixes
of \prettyref{tab:vouscalefactors}.
\end{enumerate}
The first is both unavoidable in specification, and largely
unavoidable in practice; the others are VOUnit extensions which a data
provider may of course decline to take advantage of.

The scale-factor and quoted-units extensions are intended to support
the case where the data provider wishes to distribute data including a
unit which is `unknown', but which the provider nonetheless feels is
necessary or useful; this should be done only after weighing the
considerations of Sects.~\ref{sec:scalefactor} and~\ref{sec:quoting}.
For the sake of consistency, and in order to allow
constructions such as \texttt{M'jupiterMass'}, the grammar permits quoted
units to take scaling prefixes; this is not often likely to be a good idea.

A VOUnits string which avoids the three extensions above will be
parsable, with the same meaning, in the CDS and FITS syntaxes, and
will be parsable by an OGIP parser if dots are replaced by stars.

\subsubsection{Restrictions to ASCII}

As described above, VOUnit unit strings are restricted to printable
ASCII characters.  While the two most prominent uses of these strings
will be within VOTable attributes (\verb|unit="..."|) and in XML
serialisations of a data model (for example \verb|<unit>...</unit>|),
we also intend them to be usable within FITS files and within
databases.  Neither of the latter two contexts is necessarily
unicode-friendly, so permitting non-ASCII characters in a unit string
(such as \AA\ or $\micro$) is more likely than not to cause trouble.

Similarly, forbidding spaces within VOUnit strings removes one (minor)
complication when recognising them in use.

\subsubsection{Other units, and unit-like expressions}

As noted above, the VOUnits syntax does not include structures such as
arrays or tuples of numbers.  We include in this category sexagesimal
coordinates, calendar dates (in ISO-8601 form or otherwise),
RA-Dec pairs, and other structured quantities serialised as strings.
Each of these is well-specified elsewhere, and would require a
separate parser if encountered in data.

Existing VO standards already recommend that coordinates be expressed
in decimal degrees.

Quantities like the Modified Julian Date (MJD) are also not recognized
VOUnits. As described in \prettyref{sec:notations}, the quantity MJD
can be seen as a concept (described by the appropriate UCD or utype),
and the corresponding value will most likely be expressed in days, so
the VOUnit will be \unit{d}. There is no need to overload VOUnits to
incorporate the description of concepts themselves.

The notion of unit conversion and quantity manipulation is discussed in
\prettyref{sec:conversion}.

\vskip 0pt plus 1cm
\section{Use cases and applications (informative)\label{sec:useCase}}

\subsection{Unit parsing}

The rules defined in \prettyref{sec:proposal} allow us to build VOUnit parsers.
Several services can be built on top of a VOUnit parser:

\begin{enumerate}
\item Validation. A service checking that a VOUnit is well written. The output
of such a service can have different levels: fully valid unit; valid syntax, but
not the preferred one (\eg  use of deprecated symbols); parsing error. 
\item Explanation. A service returning a plain-text explanation of the unit label.
\item Typesetting. A service returning an equivalent of the unit label suitable for inclusion in
a \LaTeX\ or HTML document.
\item Dimensional equation. As described by \citet{osuna05}, VOUnits can be translated
into a dimensional equation, allowing to build up conversions methods from one string 
representation to another one (see also \prettyref{sec:conversion}). 
\end{enumerate}

\subsection{Libraries\label{sec:libraries}}

There are a few existing libraries able to interpret unit labels.
In all cases,
some software effort is required if they are to be used in translating
between data provider unit labels, and those to be adopted by
the IVOA for internal use.

One of the most widely-used specialised
astronomical libraries is AST which includes a unit conversion
facility attached to astronomical coordinate systems \citep{berry12}.

Another library has been developed at
CDS,\footnote{\url{http://cds.u-strasbg.fr/resources/doku.php?id=units}}
and can be tested online.\footnote{\url{http://cdsweb.u-strasbg.fr/cgi-bin/Unit}} This library covers all
the symbols and notations defined in the standard for astronomical catalogues \citep[\S3.2]{cds00}, as well as
additional symbols and notations.

The Unity library\footnote{\url{https://bitbucket.org/nxg/unity}} is a new
standalone library intended to parse unit strings in the VOUnits,
OGIP, StdCats and FITS syntaxes; it was used as a vehicle for
developing and testing the grammars and 
ideas for this present document.  It provides yacc-style grammars for
the various syntaxes, as well as implementing them in parsers written
in Java and~C.  The grammars of \prettyref{appx:grammar} are extracted
from the Unity distribution.

\subsection{Unit conversion and quantity transformation\label{sec:conversion}}

Unit conversion is the simple task of converting a quantity expressed
in a given unit into a different unit, while the concept remains the
same. For example, such a library might be able to convert a distance
in \unit{pc} into a distance in \unit{AU} or \unit{km}, or convert a
flux from \unit{mJy} to \unit{W.m-2.Hz-1}. This is rather easy with
existing libraries, using dimensional analysis or SI units as a
reference.

Quantity transformation consists in deriving a new quantity from one or several original
quantities. It is more complex, because it requires having a precise model 
(a simple equation in simple cases) for computing the transformation. The model involves
quantities, each described with a UCD or utype, value and VOUnit. Some of the quantities
involved might be physical constants (\eg  Boltzmann's constant $k_{\mathrm{B}}$).

Examples of such transformations can be:
\begin{itemize}
\item linear unit conversion: a distance is measured in \unit{pixel} in an image, and needs to be transformed in
the corresponding angular separation in \unit{arcsec}. This can be done if the quantity representing the pixel
scale is given, with its value and a compatible unit like \unit{deg/pixel}.
\item converting a photon wavelength in the corresponding photon energy or frequency.
\item deriving the flux for a given photon emission rate (in \units* W) from Planck's
constant ($6.63 \times 10^{-34}\units{J~s}$), the radiation frequency (in \units{GHz}), and the
number of photons emitted per second.
\item transforming a magnitude into a flux, as needed for SED building.
\end{itemize}

VOUnits can help in quantity transformation if all quantities are qualified with proper VOUnits.

\subsection{Query languages}

Including VOUnits in queries is not an easy task. Some guidelines were
articulated during the development of the ADQL standard.

\begin{enumerate}
\item All data providers should be encouraged to supply units for each column
of a table. Columns should also have associated UCDs, so that quantities can be
properly identified.

\item The IVOA needs to provide a parser to relate the native units to the standard IVOA
labels (in this context, the `native units' are the units of the
underlying database table or metadata).   

\item
The default response to a query which does not specify units, will be
in the native units of the table. 

\item
Where queries involve combining or otherwise operating on the content
of columns to produce an output column with modified units, we can
provide libraries and a parser to assist in assigning and checking a
new unit, and attach this to the returned values via the SQL CAST
operator. 
This is implemented already in database related applications such as 
Saada,\footnote{\url{http://saada.unistra.fr/}} for instance. 
If any column used in responding to a query lacks a necessary unit, the output
involving that column will be unitless.

\item
If the user wants to change the output units with respect to the table
units, this could be done by specifying the units in the initial
SELECT statement. There are several issues to consider: 
	\begin{enumerate}
	\item Does the user also need to include the conversion expression, or does the unit
parser take care of that?  
	\item Can the user use this to assign units (based on prior knowledge) to output from a 
column lacking a unit?
	\end{enumerate}
\end{enumerate}

\subsection{Broader use in the VO}

\begin{figure}[thb]
  \includegraphics[width=\textwidth]{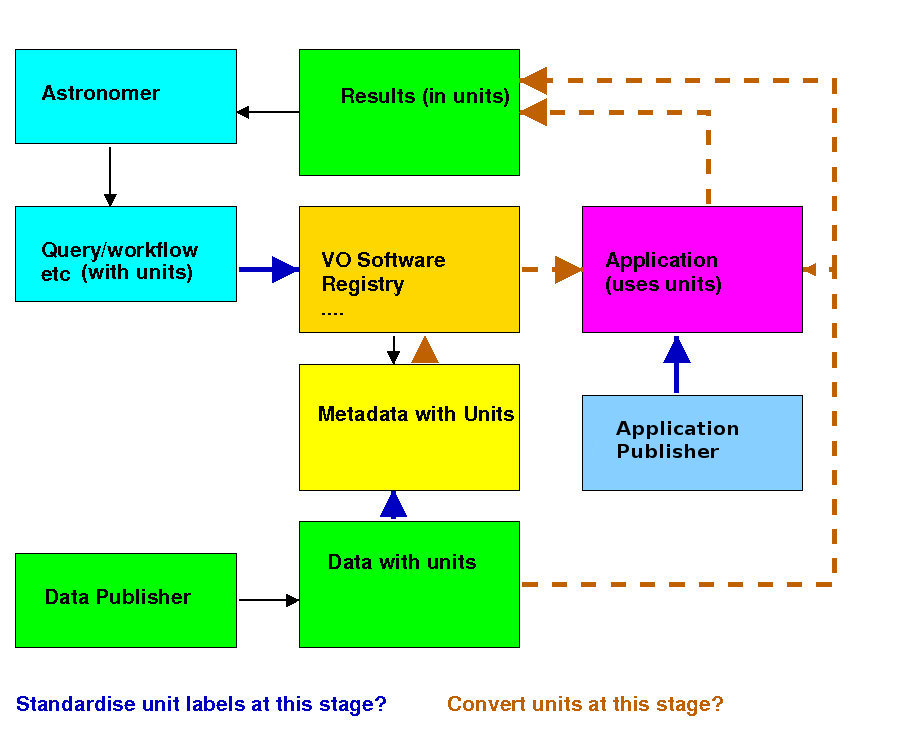}
  \caption{This shows the levels at which conversions might be done.
\textcolor{blue}{Plain arrows}: At the point where an astronomer or
  data provider submits input to the VO, we should provide tools to
  ensure that units are labeled consistently according to VOUnits. 
  This implies that a units parsing step is included prior to metadata ingestion into the VO.
\brown{Dashed arrows}: Conversions required to supply results to
  the user in specified or user-prefered units \eg  \texttt{J.s-1} to \texttt{W}, are done where and when they are required.}
  \label{fig:units2}
\end{figure}

Different VO entities require and consume metadata with units attached like registries, 
applications and interoperate via protocols. \prettyref{fig:units2} illustrates the places where the IVOA
could intervene to ensure consistent use of units.

\clearpage

\appendix

\section{Current use of units (informative)}
\label{appx:current}

Many other projects have already produced lists of preferred
representations of units. Those most commonly used in
astronomy are described in this section. 

The four first schemes described below are used as references for the
comparison tables presented later in this document.

\subsection{IAU 1989\label{appx:IAU}}

In the section 5.1 of its Style Manual, the IAU gives a set
of recommendations for representing units in publications \citep{wilkins89}. This document
therefore provides useful reference guidelines, but is not directly
applicable to VOUnits because the recommendations are more intended
for correct typesetting in journals than for standardized metadata exchange.
The IAU style will be summarized in the second column of the comparison tables.

\subsection{OGIP 1993}

NASA has defined a list of character strings specifying the basic physical units 
used within OGIP (Office of Guest Investigator Programs) FITS files \citep{george95}. Rules and guidelines on the construction 
of compound units are also outlined. 

HEASARC datasets follow these conventions, presented in the third column
of the comparison tables.

\subsection{Standards for astronomical catalogues}

The conventions adopted at CDS are summarized in the Standards for Astronomical 
Catalogues, Version 2.0 \citep[\S3.2]{cds00}. They are presented in the fourth column
of the comparison tables.

\subsection{FITS 2010}

In Section 4.3 of the reference FITS paper, \citet{pence10} describe how unit strings are to be expressed in
FITS files. The recommendations are presented in the fifth column
of the comparison tables.

\subsection{Other usages}

\begin{description}
\item[\url{http://arxiv.org/pdf/astro-ph/0511616}]
Dimensional Analysis applied to spectrum handling in VO context~\citep{osuna05}
offers a mathematical framework to guess and recompute
SI units for any quantity in astronomy.

\item[\url{http://unitsml.nist.gov}]
The NIST (National Institute of Standards \& Technology) project
UnitsML builds up an XML representation of units at the granularity
level of a simple symbol string.

\item[\url{https://www.jcp.org/en/jsr/detail?id=275}]
JAVA JSR-275 specifies Java packages for the programmatic
handling of physical quantities and their expression as numbers of
units.
\item[\texttt{aips++} and \texttt{casacore}]
These systems (see \url{http://aips2.nrao.edu/docs/aips++.html} and
 \url{http://code.google.com/p/casacore/}) contain modules handling
 units and quantities with high precision. The packages are mainly in use for
radio astronomy but are designed to be modular and adaptable (NB:
contrary to the statement on the casacore link, aips++ is still very much in
use as the toolkit behind the \textsc{casa} package).
\end{description}

\clearpage
\section{History: Comparison of syntaxes (informative)\label{appx:comparisons}}

In this section, we compare the pre-existing unit-string syntaxes and the
current standard described in this document.
We have included these comparisons for
more-or-less historical reasons, to try to highlight the variations
between syntaxes, and so illustrate the motivation for this
Recommendation, namely that the current practice, though it may at
first appear to have rough consensus, is disturbingly heterogeneous.

\begin{table}[ht]
  \begin{tabular*}\textwidth{@{\extracolsep{\fill}}|L{0.2\linewidth}|p{0.11\linewidth}|p{0.11\linewidth}|p{0.11\linewidth}|p{0.11\linewidth}|p{0.11\linewidth}|}

\hline
    & IAU & OGIP  & StdCats & FITS  & VOU\\\hline
    Units are strings of chars &  & YES &  & YES & YES\\\hline
    Case sensitive & YES & YES & YES & YES & YES\\\hline
    Character set &  &  & No spaces & ASCII text & ASCII printable\\\hline
\end{tabular*}
  \caption{Comparison of string representation and encoding.}
  \label{tabx:comparUnitEncoding}
\end{table}

\begin{table}[ht]
\begin{tabular*}\textwidth{@{\extracolsep{\fill}}|L{0.2\linewidth}|L{0.11\linewidth}|L{0.11\linewidth}|L{0.11\linewidth}|L{0.11\linewidth}|L{0.11\linewidth}|}
\hline
    & IAU & OGIP  & StdCats & FITS  & VOU\\\hline
    The 6+1 base & \multicolumn{5}{c|}{\unit{m, s, A, K, mol, cd}} \\
    \cline{2-6}
     SI units (use \unit{s}, not sec, for seconds) & (1) & \unit{kg} & \unit{g} & \unit{kg}, but \unit{g} allowed & \unit{g}\\
     \hline
    Dimensionless planar and solid angle
        & \multicolumn{3}{c|}{\unit{rad}, \unit{sr}}
        & \unit{rad}, \unit{sr}, \unit{deg} (2)
        & \unit{rad}, \unit{sr}\\ \hline
    Derived units & \multicolumn{5}{c|}{\unit{Hz, N, Pa, J, W, C, V,}} \\
    with symbols  & \multicolumn{5}{c|}{\unit{S, F, Wb, T, H, lm, lx}} \\
     & \unit{$\Omega$} & \unit{ohm} & \unit{Ohm} & \unit{Ohm} & \unit{Ohm}\\\hline
\end{tabular*}
  \caption[Comparison of base units]{Comparison of base units.  Notes: (1) unit is \unit{kg}, but use \unit{g} with prefixes; (2) \unit{deg} preferred for decimal angles}
  \label{tabx:comparUnitBase}
\end{table}


\begin{table}[ht]
\begin{tabular*}\textwidth{@{\extracolsep{\fill}} |L{0.2\linewidth}|L{0.11\linewidth}|L{0.11\linewidth}|L{0.11\linewidth}|L{0.11\linewidth}|L{0.11\linewidth}|}
\hline
    & IAU & OGIP & StdCats sec.~3.2.3 & FITS & VOU\\\hline
    Scale factors,   & \multicolumn{5}{c|}{\unit{d, c, m, n, p, f, a}} \\
    (multiple) & \multicolumn{5}{c|}{\unit{da, h, k, M, G, T, P, E}}   \\
    prefixes & \unit{\micro} & \multicolumn{3}{c|}{\unit{u}} & \unit{u}\\
     &  & \multicolumn{3}{c|}{\unit{z, y, Z, Y}} & \unit{z, y, Z, Y}\\\hline
    Prefix--symbol concatenation & (1) & (2) & no space & no space (implicit) & no space\\\hline
    Prefix-able symbols  & Not \unit{kg}: use \unit{g} & (3) & all & all & (4) \\\hline
    Use compound prefixes & should not & should never & must not & must not & must not\\\hline
\end{tabular*}
  \caption[Comparison of scale-factors]{Comparison of scale-factors.
  Notes: (1) no space, regarded as single symbol;
  (2)~no space, regarded as a single unit string;
  (3)~all units above, and \unit{eV, pc, Jy, Crab} Only \unit{mCrab} allowed;
  (4)~all (except \unit{P} for \unit{a}).}
  \label{tabx:comparUnitScale}
\end{table}

\begin{table}[ht]
\begin{tabular*}{\textwidth}{@{\extracolsep{\fill}}|L{0.2\linewidth}|L{0.11\linewidth}|L{0.11\linewidth}|L{0.11\linewidth}|L{0.11\linewidth}|L{0.11\linewidth}|}
\hline
    & IAU & OGIP  & StdCats & FITS  & VOU\\\hline
    minute & \unit{min, $^\mathrm{m}$} & \unit{min} & \unit{min} & \unit{min} & \unit{min}\\\hline
    hour & \unit{h, $^\mathrm{h}$} & \unit{h} & \unit{h} & \unit{h} & \unit{h}\\\hline
    day & \unit{d, $^\mathrm{d}$} & \unit{d} & \unit{d} & \unit{d} & \unit{d}\\\hline
    year & \unit{a} & \unit{yr} & \unit{a, yr} & \unit{a, yr} (1)& like FITS\\\hline
    arcsecond & \unit{${}''$} & \unit{arcsec} & \unit{arcsec} & \unit{arcsec} & \unit{arcsec}\\\hline
    arcminute & \unit{${}'$} & \unit{arcmin} & \unit{arcmin} & \unit{arcmin} & \unit{arcmin}\\\hline
    degree (angle) & \unit{$^\circ$} & \unit{deg} & \unit{deg} & \unit{deg} & \unit{deg}\\\hline
    milliarcsecond & \unit{mas} (use \unit{nrad}!) &  & \unit{mas} & \unit{mas} & \unit{mas}\\\hline
    microarcsec &  &  & \unit{uarcsec} &  & (2)\\\hline
    cycle & \unit{c, $^\mathrm{c}$} (5) &  &  &  & not used\\\hline
    astronomical unit & \unit{au} & \unit{AU} & \unit{AU} & \unit{AU} & \unit{AU}\\\hline
    parsec & \multicolumn{4}{c|}{\unit{pc}} & \unit{pc}\\\hline
    atomic mass & \unit{u} &  &  & \unit{u} & \unit{u}\\\hline
    electron volt & \multicolumn{4}{c|}{\unit{eV}} & \unit{eV}\\\hline
    jansky & \multicolumn{4}{c|}{\unit{Jy}} & \unit{Jy}\\\hline
    celsius degree & \unit{$^\circ$C} for meteorology, otherwise \unit{K}&  &  &  & not used\\\hline
    century & (3)&  &  &  & (4)\\\hline
\end{tabular*}
  \caption[Comparison of astronomy-related units]{Comparison of astronomy-related units.
  Notes: (1) Pa (peta-a) forbidden;
  (2) no dedicated symbol, use \unit{uarcsec};
  (3) ha, cy should not be used;
  (4) no dedicated symbol, use \unit{ha} or \unit{hyr};
  (5) superscript-`c' has also been used to denote `radian'
  \label{tabx:comparUnitAstro}}
\end{table}

\begin{table}[ht]
\begin{tabular*}{\textwidth}{@{\extracolsep{\fill}}|L{0.2\linewidth}|L{0.11\linewidth}|L{0.11\linewidth}|L{0.11\linewidth}|L{0.11\linewidth}|L{0.12\linewidth}|}
\hline
    & IAU & OGIP  & StdCats & FITS  & VOU\\\hline
    \aa{}ngstr\"om &\unit{\AA}& \unit{angstrom} & 0.1nm & \unit{Angstrom} & \unit{angstrom}, \unit{Angstrom}\\\hline
    micron & \unit{\micro} &  &  &  & not used \\\hline
    fermi & no symbol &  &  &  & not used \\\hline
    barn & \unit{b} & \unit{barn} & \unit{barn} & \unit{barn} & \unit{barn}\\\hline
    cubic centimetre & \unit{cc} &  &  &  & no dedicated symbol\\\hline
    dyne & \unit{dyn} & \unit{} & \unit{} & \unit{} & not used \\\hline
    erg & \unit{erg} & \unit{erg} & (1) & \unit{erg} & \unit{erg} \\\hline
    calorie & \unit{cal} & \unit{} & \unit{} & \unit{} & not used \\\hline
    bar & \unit{bar} & \unit{} & \unit{} & \unit{} & not used \\\hline
    atmosphere & \unit{atm} & \unit{} & \unit{} & \unit{} & not used \\\hline
    gal & \unit{Gal} & \unit{} & \unit{} & \unit{} & not used \\\hline
    eotvos & \unit{E} & \unit{} & \unit{} & \unit{} & not used \\\hline
    gauss & \unit{G} & \unit{G} & \unit{} & \unit{G} & \unit{G} \\\hline
    gamma & \unit{$\gamma$} & \unit{} & \unit{} & \unit{} & not used \\\hline
    oersted & \unit{Oe} & \unit{} & \unit{} & \unit{} & not used \\\hline
    Imperial, non-metric & should not be used & \unit{} & \unit{} & \unit{} & not used \\\hline
\end{tabular*}
  \caption[Comparison of symbols deprecated by IAU]{Comparison of
  symbols deprecated by IAU (from \citet{wilkins89} table~7, ``Non-SI
  units and symbols whose continued use is deprecated'').
  Note: (1) no symbol: \unit{mW/m2} is used for \units{erg\,cm^{-2}\,s^{-1}}.}
  \label{tabx:comparUnitDeprecated}
\end{table}

\begin{table}[ht]
\begin{tabular*}{\textwidth}{@{\extracolsep{\fill}}|L{0.2\linewidth}|L{0.11\linewidth}|L{0.11\linewidth}|L{0.11\linewidth}|L{0.11\linewidth}|L{0.11\linewidth}|}
\hline
    & IAU & OGIP  & StdCats & FITS  & VOU\\\hline
    magnitude & \multicolumn{4}{c|}{\unit{mag}} & \unit{mag}\\\hline
    rydberg & \unit{} & \unit{} & \unit{Ry} & \unit{Ry} & \multirow{19}{0.15\linewidth}{same as FITS} \\\hline
    solar mass & \unit{$\mathrm{M}_\odot$} &  & \unit{solMass} & \unit{solMass} &\\\cline{1-5}
    solar luminosity & \unit{} & \unit{} & \unit{solLum} & \unit{solLum} &\\\cline{1-5}
    solar radius & \unit{} & \unit{} & \unit{solRad} & \unit{solRad} &\\\cline{1-5}
    light year & \unit{} & \unit{lyr} & \unit{} & \unit{lyr} &\\\cline{1-5}
    count & \unit{} & \unit{count} & \unit{ct} & \unit{ct, count} &\\\cline{1-5}
    photon & \unit{} & \unit{photon} & \unit{} & \unit{photon, ph} &\\\cline{1-5}
    rayleigh & \unit{} & \unit{} & \unit{} & \unit{R} &\\\cline{1-5}
    pixel & \unit{} & \unit{pixel} & \unit{pix} & \unit{pix, pixel} &\\\cline{1-5}
    debye & \unit{} & \unit{} & \unit{D} & \unit{D} &\\\cline{1-5}
    relative to Sun & \unit{} & \unit{} & \unit{Sun} & \unit{Sun} &\\\cline{1-5}
    channel & \unit{} & \unit{chan} & \unit{} & \unit{chan} &\\\cline{1-5}
    bin & \unit{} & \unit{bin} & \unit{} & \unit{bin} &\\\cline{1-5}
    voxel & \unit{} & \unit{voxel} & \unit{} & \unit{voxel} &\\\cline{1-5}
    bit & \unit{} & \unit{} & \unit{bit} & \unit{bit} &\\\cline{1-5}
    byte & \unit{} & \unit{byte} & \unit{byte} & \unit{byte} &\\\cline{1-5}
    adu & \unit{} & \unit{} & \unit{} & \unit{adu} &\\\cline{1-5}
    beam & \unit{} & \unit{} & \unit{} & \unit{beam} &\\\hline
     & \unit{} & \unit{Crab} avoid use & \unit{} & \unit{} & not used \\\hline
    No unit, dimensionless & \unit{} & blank string & \unit{-} & \unit{} & empty string \\\hline
    Percent &  &  & \unit{\%} & & \unit{\%} \\\hline
    unknown & \unit{} & {\tiny\unit{UNKNOWN}} & \unit{} & \unit{} & \unit{unknown} \\\hline
\end{tabular*}
  \caption{Miscellaneous other symbols.}
  \label{tabx:comparUnitOther}
\end{table}

\begin{table}[th]
\begingroup
\begin{tabular*}{\textwidth}{@{\extracolsep{\fill}}|L{0.2\linewidth}|L{0.11\linewidth}|L{0.11\linewidth}|L{0.11\linewidth}|L{0.11\linewidth}|L{0.11\linewidth}|}
\hline
    & IAU & OGIP  & StdCats & FITS &VOU\\\hline
    Multiplication
        & space or dot (1)
    	& space or star (2)
	& dot
	& space, dot or star (3)
        & dot \\\hline
    Division
        & per (4)
    	& \unit{/} (5)
	& \unit{/}, no space
	& \unit{/}, no space
        & \unit{/}, no space\\\hline
    Use of multiple /
        & never
    	& allowed 
	& allowed 
	& discouraged (6)
        & never\\\hline
    \unit{sym} raised to the power $y$
        & superscript 
    	& (7)
	& (8)
	& (9)
        & \unit{**}\\\hline
    Exponential of \unit{sym}
        &
        & \unit{exp(sym)}
        &
        & \unit{exp(sym)}
        & \unit{exp(sym)}\\\hline
    Natural log of \unit{sym}
        &
        & \unit{ln(sym)}
        &
        & \unit{ln(sym)} 
        & \unit{ln(sym)} \\\hline
    Decimal log of \unit{sym}
        &
        & \unit{log(sym)}
        & \unit{[sym]}
        & \unit{log(sym)}
        & \unit{log(sym)} \\\hline
    Square root of \unit{sym}
        &
        & \unit{sqrt(sym)}
        &
        & \unit{sqrt(sym)}
        & \unit{sqrt(sym)} \\\hline
    Other math &  & (10) &  & not used  & not used \\\hline
    ( ) &  & allowed & allowed & optional around powers & allowed \\\hline
    powers & superscripts & (11) & integers & (12) & (12) \\\hline
    Numeric factor & not used & (13) & allowed & (14) & (15) \\\hline
\end{tabular*}
\endgroup
\caption[Mathematical expressions and combinations]{Mathematical
    expressions and symbol combinations.
    This table is derived from the specification texts; any deviation
    from the grammars
    of \prettyref{appx:fitsgrammar}--\ref{appx:cdsgrammar} is
    unintentional.  This table, and those appendices, are informative;
    only \prettyref{appx:vougrammar} is normative.
  \label{tabx:comparUnitCombine}
  Notes: (1) space, except if previous unit ends with superscript; dot (\unit{.}) may be used;
  (2)~one or more spaces OR one asterisk (\unit{*}) with optional spaces on either side;
  (3)~single space OR asterisk (\unit{*}, no spaces) OR dot (\unit{.}, no spaces);
  (4)~use negative index or solidus (\unit{/});
  (5)~solidus (\unit{/}) with optional spaces on either side, space not recommended after / OR negative index;
  (6)~may be used, but discouraged, `math precedence rule';
  (7)~\unit{sym**($y$)} parenthesis optional if $y>0$;
  (8)~nothing -- \unit{sym$y$}, and use $+/-$ sign for \unit{10+21};
  (9)~\unit{sym$y$} OR \unit{sym**($y$)} OR \unit{sym\^{}($y$)}, no space;
  (10)~\unit{$f$(sym)}, where $f$ is
\unit{sin}, \unit{cos}, \unit{tan}, \unit{asin}, \unit{acos}, \unit{atan}, \unit{sinh}, \unit{cosh}, \unit{tanh};
  (11)~decimal and integer fractions allowed;
  (12)~integer (sign and () optional), OR decimal or ratio between ();
  (13)~should be avoided; only powers of 10 allowed; should precede any unit string;
  (14)~optional 10**k, 10\texttt{\^}k, or 10$\pm$k;
  (15)~see \prettyref{sec:scalefactor}.}
\end{table}

\clearpage
\section{Formal grammars\label{appx:grammar}}

\emph{Subsection \ref{appx:vougrammar} is Normative, the other
    subsections are Informative.}

In this section we provide formal (yacc-style) grammars for the four
ASCII-based syntaxes discussed in this document.  The FITS, OGIP and
CDS grammars are not normative: the corresponding specification
documents do not provide grammars, and instead describe the syntaxes
in text, so that the grammars here are deductions from the
specification text.
This unfortunately means that some of these syntaxes were discovered to be ambiguous.
These ambiguities are discussed in the sections below.  We recommend
that VO applications parse these syntaxes in a way which is consistent
with the grammars here.
The grammar for the VOUnits syntax, in \prettyref{appx:vougrammar}, is normative.

We believe that the grammars below are such that if a string 
successfully parses in two distinct grammars, it means the same in
both.

The grammars here are from release 1.0 of the `Unity' package at
\url{https://bitbucket.org/nxg/unity}, which includes machine-readable
grammars, lists of recommended units, and a collection of test cases.  These are also extracted in
machine-readable form
at \url{https://code.google.com/p/volute/source/browse/trunk/projects/std-vounits/unity-grammars.zip}.

In these grammars, the common terminals are as given in
\prettyref{tabx:terminals}.  Lexers \norm{must not} swallow whitespace
in generating these terminals; whitespace is permitted in a units
string only where the corresponding grammar permits
the \texttt{WHITESPACE} terminal.

\begin{table}[ht]
\begin{tabular}{rL{9cm}}
\texttt{CARET}&the \texttt{\^{}} character (\hex{5e})\\
\texttt{DIVISION}&the solidus, \texttt{/} (\hex{2f})\\
\texttt{DOT}&the dot/period/full-stop character (\hex{2e})\\
\texttt{FLOAT}&a string matching the regular expression
       \texttt{[-+]?[0-9]+\textbackslash.[0-9]+}\\
\texttt{LIT10}&a literal string `\texttt{10}' (the sequence \hex{31} \hex{30})\\
\texttt{OPEN\_P} / \texttt{CLOSE\_P}&parentheses (\hex{28} and \hex{29})\\
\texttt{SIGNED\_INTEGER}&an integer with a required leading sign, so
matching the regular expression \texttt{[-+][0-9]+}\\
\texttt{STAR}&the asterisk (\hex{2a})\\
\texttt{STARSTAR}&a pair of asterisks, \texttt{**}\\
\texttt{STRING}&a non-empty sequence of letters \texttt{[a-zA-Z]+}\\
\texttt{UNSIGNED\_INTEGER}&an integer with no leading sign \texttt{[0-9]+}\\
\texttt{WHITESPACE}&a non-empty string of space characters (\hex{20} only)\\
\end{tabular}
\caption[The terminals used in the grammars]
{\label{tabx:terminals}The terminals used in the grammars; the
notation \hex{nn} indicates hexadecimal ASCII character numbers;
the digits are \hex{30} to \hex{39}, the letters are \hex{41} to \hex{5a} and \hex{61} to
\hex{7a}, and the sign characters are \hex{2b} and \hex{2d}.}
\end{table}

\subsection{The FITS grammar (informative)}
\label{appx:fitsgrammar}

For the FITS units syntax, see section~4.3 of~\cite{pence10}, and its
associated tables.  Our preferred FITS grammar is in
\prettyref{tabx:fitsgrammar}.

As noted above in \prettyref{sec:fitsquote}, 
the FITS specification isn't completely clear on the topic of 
solidi, saying ``[t]he IAU style manual forbids
the use of more than one solidus (/) character in a units
string. However, since normal mathematical precedence rules apply
in this context, more than one solidus may be used but is
discouraged''.  This does not really resolve the question of whether, for
example, \texttt{kg/m s} should be parsed as \units{kg~m^{-1}~s^{-1}}
or as \units{kg~m^{-1}~s}, since this is a question of both operator
precedence and (left-)associativity, where there might be different
rules internationally, and conflicts between mathematical and
programming-language rules.  Most people would \emph{probably} parse
it as \units{kg~m^{-1}~s^{-1}}, but we trust that most educators would
oblige students to rewrite the expression on the grounds that any
ambiguity is too much.
Here, we resolve the ambiguity by declaring that there can
be only a single expression to the right of the solidus.

It is a consequence of this that nothing can be
successully parsed in two different grammars, with different
meanings.  If the right-hand-side of the division could be a
\texttt{product\_of\_units}, then \texttt{kg /m s} would parse in both
the FITS and OGIP syntaxes,
but mean \units{kg~m^{-1}~s^{-1}} in the FITS syntax, and
\units{kg~m^{-1}~s} in the OGIP one.

The FITS specification permits a leading numeric multiplier, but
``[c]reators of FITS files are encouraged to use the numeric
multiplier only when the available standard scale-factors of [SI] will
not suffice''.

The FITS specification permits \texttt{m(2)}, to indicate the square of
unit~`m'.  The grammar has to special-case this, in order to
distinguish it from function application.

Other ambiguities:
\begin{itemize}
\item The FITS specification may or may not be intended to permit 
  \texttt{10+3 /m}, but we don't.
\item It is possible to read the FITS spec as permitting
  \texttt{m\^{}1.5}, without parentheses.  We take it to be
  invalid here.
\end{itemize}

\clearpage
\begin{table}[t]
\verbatiminput{unity-grammars/unity-fits.txt}
\caption[The FITS grammar]{\label{tabx:fitsgrammar}The FITS grammar.
See \prettyref{appx:fitsgrammar}.}
\end{table}
\clearpage

\subsection{The OGIP grammar (informative)}
\label{appx:ogipgrammar}

For the OGIP units syntax, see \cite{george95}.  Our preferred OGIP
grammar is in \prettyref{tabx:ogipgrammar}.

The OGIP specification somewhat reluctantly concedes (in its section
3.2) that ``occasionally it may be preferable to include [leading
scale] factors on the grounds of user-friendliness'', but that ``[t]he
inclusion of numerical factors should therefore be avoided wherever
possible'', and it is ``suggested'' that the scale-factor should in any case
be restricted to powers of~10.

Specification ambiguities:
\begin{itemize}
\item The OGIP specification permits a space between the leading
  factor and the rest of the unit (by implication from the provided
  examples).
\item The specification does not indicate the format of the numerical
  factor in the case where it is not a power of ten.  We have
  suggested \texttt{FLOAT} here (see \prettyref{tabx:terminals}).
\item OGIP \emph{recommends} having no whitespace after the division
  solidus, but does not forbid it; therefore we permit it in this
  grammar.
\item From its specification text, OGIP appears to permit
  \texttt{str1**y}, where \texttt{y} can be a float, even though none
  of its examples include this.  The same interpretive logic would
  appear to permit \texttt{m**3/2}, but this seems to run too great a
  risk of being misparsed, and we forbid it here.
\item In the same place, the text suggests that \texttt{str1**y} may
  omit the brackets `if~\texttt y is positive', but the context
  suggests that the intention is to permit this if~\texttt y is
  unsigned.  In the grammar here, we permit the omission of the
  brackets only if~\texttt y is unsigned -- that is, \texttt{m**+2},
  like \texttt{m**-2}, is forbidden.
\end{itemize}

\begin{table}[ht]
\verbatiminput{unity-grammars/unity-ogip.txt}
\caption[The OGIP grammar]{\label{tabx:ogipgrammar}The OGIP grammar.
Note that the \texttt{FLOAT} in the \texttt{scalefactor} production
must be a power of ten.
See \prettyref{appx:ogipgrammar}.}
\end{table}
\clearpage

\subsection{The CDS grammar (informative)}
\label{appx:cdsgrammar}

For the CDS units syntax, see \cite[\S3.2]{cds00}.  Our preferred CDS
grammar is in \prettyref{tabx:cdsgrammar}.  It requires additional
terminals, described in \prettyref{tabx:cdsterminals}.

Specification ambiguities:
\begin{itemize}
\item The CDS document indicates that units should be raised to powers by
concatenation of the unit string with an integer, but does so rather
elliptically, so that it is not clear whether \texttt{m+2} is
permitted (the relevant examples show this as \texttt{m2}).  We take
this to be permitted in this grammar.
\item The specification does not indicate the format of the numerical
  factor in the case where it is not a power of ten and not
  a \texttt{CDSFLOAT}.  We have suggested \texttt{FLOAT} here
  (see \prettyref{tabx:terminals}).
\item The document does not specify or illustrate how \texttt{kg/m/s}
should be parsed.  Since the document mentions the OGIP standard (even
though it does not permit OGIP's syntax for powers, \texttt{m**2}), we
take it that this is valid, and equivalent to \units{kg~m^{-1}~s^{-1}}.
\end{itemize}

This specification places no restrictions on the leading scale-factor.

\begin{table}[ht]
\verbatiminput{unity-grammars/unity-cds.txt}
\caption[The CDS grammar]{\label{tabx:cdsgrammar}The CDS grammar.
See \prettyref{appx:cdsgrammar} for discussion,
and \prettyref{tabx:cdsterminals} for the additional terminals.}
\end{table}
\begin{table}[ht]
\begin{tabular}{rL{10cm}}
\texttt{CDSFLOAT}&a string matching the regular
expression \texttt{[0-9]+\textbackslash.[0-9]+x10[-+][0-9]+}
(that is, something resembling \texttt{1.5x10+11})\\
\texttt{OPEN\_SQ}&the open square bracket `\texttt{[}' (indicates logs
  in this syntax)\\
\texttt{CLOSE\_SQ}&the close square bracket `\texttt{]}'\\
\texttt{PERCENT}&the percent character `\%'
\end{tabular}
\caption[Extra CDS terminals]{\label{tabx:cdsterminals}Extra terminals
for the CDS grammar}
\end{table}
\clearpage

\subsection{The VOUnits grammar (normative)}
\label{appx:vougrammar}

The VOUnits grammar is defined by this section,
by the grammar in \prettyref{tabx:vougrammar}
(with the terminals of \prettyref{tabx:terminals}
plus the extra terminals listed in \prettyref{tabx:vounitsterminals}),
the list of known units of \prettyref{tabx:knownunits},
and the list of known functions of \prettyref{tab:functions}.

The intention of the VOUnits grammar is that if a VOUnits string
does not use the scale-factor, quoted-units or binary-prefix
extensions
(that is, if it avoids the \texttt{VOUFLOAT}
and \texttt{QUOTED\_STRING} terminals and is restricted to SI decimal prefixes),
then it will be parsable, with the same semantics, by FITS
and CDS parsers, and that it will be parsable by an OGIP parser if
dots are replaced by stars.
See \prettyref{sec:deviations} for discussion.
In particular:
\begin{itemize}
\item The product of units is indicated only by a dot, with no
  whitespace: \texttt{N.m}.
\item Raising a unit to a power is done only with a double-star:
  \texttt{kg.m**2.s**-2}.
\item There may be at most one division sign at the top level of an
  expression.
\end{itemize}

In \prettyref{tabx:vougrammar}, the \texttt{VOUFLOAT} terminal is a
string matching either of the regular expressions
\begin{itemize}
\item\texttt{0\textbackslash.[0-9]+([eE][+-]?[0-9]+)?}
\item\texttt{[1-9][0-9]*(\textbackslash.[0-9]+)?([eE][+-]?[0-9]+)?}
\end{itemize}
(that is, something resembling \texttt{0.123} or \texttt{1.5e+11}).

\begin{table}[ht]
\verbatiminput{unity-grammars/unity-vounits.txt}
\caption[The VOUnits grammar]{\label{tabx:vougrammar}The VOUnits
grammar.  See \prettyref{appx:vougrammar} for discussion,
and \prettyref{tabx:vounitsterminals} for additional terminals.}
\end{table}
\begin{table}[ht]
\begin{tabular}{rL{10cm}}
\texttt{VOUFLOAT}&see text, \prettyref{appx:vougrammar}\\
\texttt{QUOTED\_STRING}&a \texttt{STRING} between single quote marks
    (ASCII \hex{27})
\end{tabular}
\caption[Extra VOUnits terminals]{\label{tabx:vounitsterminals}Extra terminals
for the VOUnits grammar}
\end{table}
\clearpage

\section{Updates of this document (informative)}

The detailed (line-by-line) history of the document can be found at 
\url{https://code.google.com/p/volute/source/browse/trunk/projects/std-vounits/VOUnits.tex}.
\begin{itemize}
\item 2014 July 2: minor adjustment, to remove `proposed
recommendation' text which was retained in the final version by accident.
\item REC-1.0-20140523: Approved as REC by Exec, at the IVOA Madrid Interop.
\item PR-1.0-20140513:
A few rewordings, addressing comments made during the TCG review period.
\item PR-1.0-20140226:
Minor wording and layout changes, following on-list discussion.
Released for TCG review.
\item PR-1.0-20131224:
\begin{itemize}
\item Grammar changes: minor (now incorporates the grammars of Unity v0.11).
\item Various clarifications to the text, following on-list discussion.
\end{itemize}
\item WD-1.0-20131025:
\begin{itemize}
\item Grammar changes: The `\%' character is now treated as a special
    case, rather than being a permitted 'STRING' character; it's only
    the CDS syntax that permits this character.  Some readability
    adjustments to the grammars.  Unit strings with leading slashes
    (eg \unit{/m3}) are no longer supported in the VOUnits syntax.
    The grammars now match Unity v0.10.
\item Changed discussion/rationale for forbidding non-ASCII
    characters.
\item Clarified that `?' -- which is specified as indicating an
    unknown unit -- is not part of the VOUnits grammar, and should be
    spotted by a caller before parsing begins.
\item Clarified the extra terminals which some grammars use.
\item Clarified that the ambiguity in \unit{dadu} should remain
    unresolved, and the correct behaviour unspecified (is it
    deci-\texttt{adu} or deka-\texttt{du}?).
\end{itemize}
\item WD-1.0-20131011: Changed gramme in gram; removed color property to distinguish arrows in fig .2;
Removed astro'l unit abbreviation from known-units.tex
\item WD-1.0-20130922: Responding to RFC and mailing list comments.
Addition of quoted units and arbitrary scale-factor (so updates to
grammars, which now match Unity v0.9).  Some reformatting of tables.
\item WD-1.0-20130724: Rephrasing and clarification, responding to RFC
comments.  Update unity grammars to current version (ie, version of 2013-07-22 18:40).
\item WD-1.0-20130701: Simplified Architecture diagram. Added example
with scientific notation.  Adjusted locations of grammar tables to try
to keep them closer to the associated text.
\item WD-1.0-20130429: Some restructuring, some rephrasing, and a few layout changes.
\item WD-1.0-20130225: Large tables from section 3 moved to Appendix A. Short summaries of symbols added
to section 3. Changes to table of known units for consistency with text. Added explanations for units Sun and byte.  
\item WD-1.0-20121212:
Minor typographical fixes. Added definition of OGIP. Removed last sentence from acknowledgements, which have been moved to the beginning of the document. Changed figure 1 to move Units in Semantics. Added 'discouraged' in first line of \prettyref{tab:VOUnitCombine}. Color change in figure 2 and its label.
\item WD-1.0-20120801:
Minor typographical fixes
\item WD-1.0-20120801:
  \begin{itemize}
    \item Included yacc-style grammars in document.
    \end{itemize}
\item WD-1.0-20120718:
	\begin{itemize}
	\item Removed external tables refs in tables to avoid confusion.
	\item Removed refs to SOFA and NOVAS.
	\item Precision on the "no unit" case in text.
	\item Added formal grammar in annex.
	\item Minor editing and typo fixes.
	\end{itemize}
\item WD-1.0-20120521:
	\begin{itemize}
	\item Typos fixed, removed F. Bonnarel from authors. 
	\item One sentence rephrased in section 1.2 for clarity.
	\item Clarification of \unit{g} and \unit{kg} issue in \prettyref{sec:baseUnits}.
	\item Added remark on \unit{Pa} in \prettyref{sec:scaleFactors}.
	\item Micro-arcsecond and century explained in \prettyref{tabx:comparUnitAstro}.
	\item \prettyref{tabx:comparUnitDeprecated} completed.
	\item Added numeric factors in \prettyref{tabx:comparUnitCombine} and discussion in text.
	\end{itemize}
\item WD-1.0-20111216: Major rework of the document.
\item 0.3: initial public release.
\end{itemize}

\clearpage
\bibliographystyle{plainnat-eprints}
\bibliography{bib}

\end{document}